\documentclass[prl,twocolumn]{revtex4}

\usepackage{color,graphicx}
\usepackage{dcolumn}
\usepackage{bm}

\def\be{\begin{equation}}
 \def \ee{\end{equation}}
\def\bea{\begin{eqnarray}}
  \def\eea{\end{eqnarray}}

\begin{document}

\title{
\vspace*{-1cm}
\begin{flushright}
{{\small \sl Version 23(\today)}}
\end{flushright}
\vspace*{-0.3cm}
A Transactional Analysis of Interaction Free Measurements}
\author{John G. Cramer}
\affiliation{Department of Physics, University of Washington\\
  Seattle, WA 98195-1560}

\date{\today}

\begin{abstract}
The transactional interpretation of quantum mechanics is applied to the ``interaction-free" measurement scenario of Elitzur and Vaidman and to the Quantum Zeno Effect version of the measurement scenario by Kwiat, et al.  It is shown that the non-classical information provided by the measurement scheme is supplied by the probing of the intervening object by incomplete offer and confirmation waves that do not form complete transactions or lead to real interactions.
\end{abstract}
\maketitle

\vskip 0.5in
\section*{Introduction}

The transactional interpretation of quantum mechanics\cite{cramer-86} is an explicitly nonlocal and relativistically invariant alternative to the Copenhagen interpretation.  It requires a ``handshake" between retarded waves ($\psi$) and advanced waves ($\psi *$) for each quantum event or ``transaction" in which energy, momentum, angular momentum, and other conserved quantities are transferred.  The transactional interpretation offers the advantages over its alternatives that (1) it is actually ``visible" in the formalism of quantum mechanics, (2) it is economical, involving fewer independent assumptions than its rivals, (3) it is paradox-free, resolving all of the paradoxes and counter-intuitive aspects of standard quantum theory including nonlocality and wave function collapse, (4) it does not give a privileged role to observers or measurements, and (5) it permits the visualization of quantum events.

In the original 1986 publication of the transactional interpretation\cite{cramer-86}, all of the then-prevailing quantum paradoxes were analyzed, including Renninger's Paradox, EPR, Schr\"{o}dinger's Cat, and Wheeler's Delayed Choice.  In this paper, we will apply the transactional interpretation to the ``interaction-free" measurement schemes, devised since the transactional interpretation papers were published, which are demonstrations of another non-classical quantum effect.

\section*{The Elitzur and Vaidmann Interaction-Free Measurement}

In 1993 Elitzur and Vaidmann\cite{elitzur-93} (EV) showed a surprised physics community that quantum mechanics permits the use of light to examine an object without a single photon of the light actually interacting with the object.  The EV thought-experiment requires only the {\it possibility} of an interaction.
In their paper\cite{elitzur-93} Elitzur and Vaidmann discuss their scenario in terms of the standard Copenhagen interpretation of quantum mechanics, in which the interaction-free result is rather mysterious, particularly since the measurement permits ``knowledge" that is not available classically.  They also considered their scenario in terms of the Everett-Wheeler or ``many-worlds" interpretation of quantum mechanics.  Considering the latter, they suggest that the information indicating the presence of the opaque object can be considered to come from an interaction that occurs in a separate Everett-Wheeler universe and to be transferred to our universe through the absence of interference.  In the present paper, we will examine the same scenario in terms of the transactional interpretation and will provide a more plausible account of the physical processes behind two types of interaction-free measurements.

\begin{figure}
\includegraphics[width=8 cm]{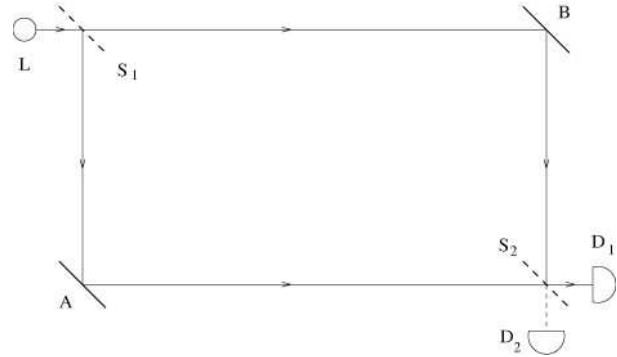}
\caption{\label{one}  Mach Zender Interferometer with both beam paths open.}
\end{figure}

We will begin by reviewing the original EV interaction-free scenario.  The basic apparatus used is a Mach-Zender interferometer, as shown in Fig. \ref{one}.  Light from a light source $L$ goes to a 50\%-50\% beam splitter $S_{1}$ that divides incoming light into two possible paths or beams.  These beams are deflected by $45^{0}$ mirrors $A$ and $B$, so that they meet at a second beam splitter $S_{2}$, which recombines them by another reflection or transmission.  The combined beams from $S_{2}$ then go to the photon detectors $D_{1}$ and $D_{2}$.

The Mach-Zender interferometer has the characteristic that, if the paths from $S_{1}$ to $S_{2}$ via $A$ and $B$ have precisely the same path lengths, the superimposed waves from the two paths are in phase at $D_{1}$ and out of phase at $D_{2}$.  This is because with beam splitters like $S_{1}$ and $S_{2}$, an emerging reflected wave is always $90^{0}$ out of phase with the corresponding transmitted wave\cite{mandel-95}.  The result is that all photons from light source $L$ will go to detector $D_{1}$ and none will go to detector $D_{2}$.

Now we place an opaque object ($Obj$) on the lower path after mirror $A$.  It will block light waves along the lower path after reflection from mirror $A$, insuring that all of the light arriving at beam splitter $S_{2}$ had traveled via path $B$.  In this case there is no interference, and the $S_{2}$ beam splitter sends equal components of the incident wave into to the two detectors.

\begin{figure}
\includegraphics[width=8 cm]{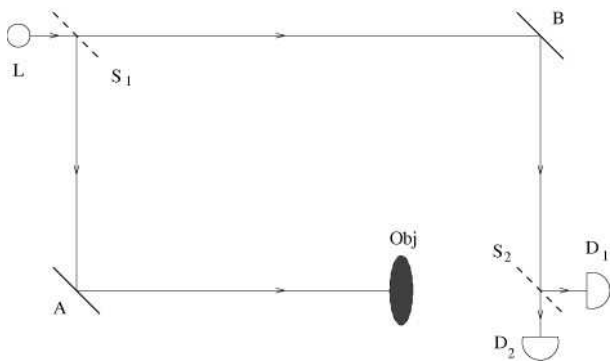}
\caption{\label{two}  Mach Zender Interferometer with one beam path blocked.}
\end{figure}

Now suppose that we arrange for the light source $L$ to emit only one photon within a given time period.  Then, if we do the measurement with no opaque object on path $A$, we should detect the photon at $D_{1}$ 100\% of the time.  If we perform the same measurement with the opaque object blocking path $A$, we should detect a photon at $D_{1}$ 25\% of the time, a photon at $D_{2}$ 25\% of the time, and should detect no photon at all 50\% of the time (because it was removed by the object in path $A$). In other words, the detection of a photon at $D_{2}$ guarantees that an opaque object is blocking path $A$, although no photon has actually interacted with object.  This is the essence of the Elitzur and Vaidmann interaction-free measurement.

Note that if a photon is detected at detector $D_{1}$, the issue of whether an object blocks path $A$ is unresolved.  However, in that case another photon can be sent into the system.  The net result of such a procedure is that 66\% of the time a photon will strike the object, resulting in no detection signal, while 33\% of the time a photon will be detected at $D_{2}$, indicating without interaction that an object blocks the $A$ path.  Thus, the EV procedure has an efficiency for non-interactive detection of 33\%.

\section*{A Transactional Interpretation of the Elitzur and Vaidmann \it{Gedankenexperiment}}

The transactional interpretation of quantum mechanics\cite{cramer-86} considers any quantum event that involves the exchange of conserved quantities (energy, momentum, angular momentum, ...) and can be  represented by a matrix element to have formed in three stages: (1) an ``offer wave" (the usual retarded wave function $\psi$ or Dirac ``ket" state vector $\mid a >$ originates from the ``source" (the object supplying the quantities transferred) and spreads through space-time until it encounters the ``absorber" (the object receiving the conserved quantities); (2) the absorber responds by producing an advanced ``confirmation wave" (the complex conjugate wave function $\psi*$ or Dirac ``bra" state vector $<a \mid$ which travels in the reverse time direction back to the source, arriving with an amplitude of $\psi\psi*$; and (3) the source chooses between the possible transactions $i$ based on the strengths of the $\psi_{i}\psi_{i}*$ echoes it receives, and reinforces the selected transaction repeatedly until the conserved quantities are transferred and the potential quantum event becomes real.  We will apply this interpretation to the quantum events of the Elitzur-Vaidmann interaction-free measurement scenario.

In what follows we will explicitly indicate offer waves by a specification of the path in a Dirac ket state vector, and we will underline the symbols for optical elements at which a reflection has occurred.  Confirmation waves will similarly be indicated by a Dirac bra state vector, and will indicate the path considered by listing the elements in the time-reversed path with reflections underlined.

\begin{figure}
\includegraphics[width=9 cm]{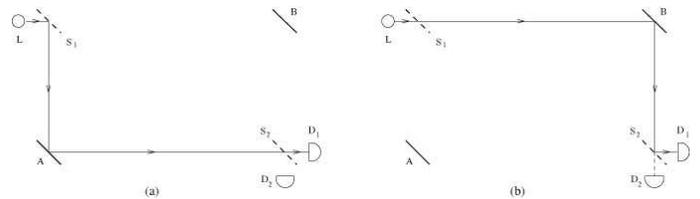}
\caption{\label{three}  Offer waves (a) $\mid L$-$\underline{S_{1}}$-$\underline{A}$-$S_{2}$-$D_{1}>$  and (b) $\mid L$-$S_{1}$-$\underline{B}$-$\underline{S_{2}}$-$D_{1}>$.}
\end{figure}

Consider first the situation in which no object is present in path $A$ as shown in Fig. \ref{three}.  The offer waves from $L$ to detector $D_{1}$ are  $\mid L$-$\underline{S_{1}}$-$\underline{A}$-$S_{2}$-$D_{1}>$ and  $\mid L$-$S_{1}$-$\underline{B}$-$\underline{S_{2}}$-$D_{1}>$.  They arrive at detector $D_{1}$ in phase because the offer waves on both paths have been transmitted once and reflected twice.  The offer wave from $L$ initially has unit amplitude, but the splits at $S_{1}$ and $S_{2}$ each reduce the wave amplitude by 1/$\sqrt{2}$, so that each wave, having been split twice, has an amplitude of $\frac{1}{2}$ as it reaches detector $D_{1}$.  Therefore, the two offer waves of equal amplitude and phase interfere constructively, reinforce, and produce a confirmation wave that is initially of unit amplitude.

Similarly, the offer waves from $L$ to detector $D_{2}$ are $\mid L$-$\underline{S_{1}}$-$\underline{A}$-$\underline{S_{2}}$-$D_{2}>$ and $\mid L$-$S_{1}$-$\underline{B}$-$S_{2}$-$D_{2}>$.  They arrive at detector $D_{2}$ $180^{0}$ out of phase, because the offer wave on path $A$ has been reflected three times while the offer wave on path $B$ has been transmitted twice and reflected once. Therefore, the two waves with amplitudes $\pm\frac{1}{2}$ interfere destructively, cancel at detector $D_{2}$, and produce no confirmation wave.

The confirmation waves from detector $D_{1}$ to $L$ are $<D_{1}$-$S_{2}$-$\underline{A}$-$\underline{S_{1}}$-$L\mid$ and $<D_{1}$-$\underline{S_{2}}$-$\underline{B}$-$S_{1}$-$L\mid$.  They arrive back at the source $L$ in phase because, as in the previous case, the confirmation waves on both paths have been transmitted once and reflected twice.  As before the splits at $S_{1}$ and $S_{2}$ each reduce the wave amplitude by $1/\sqrt{2}$, so that each confirmation wave has an amplitude of $\frac{1}{2}$ as it reaches source $L$.  Therefore, the two offer waves interfere constructively, reinforce, and have unit amplitude.  Since the source $L$ receives a unit amplitude confirmation wave from detector $D_{1}$ and no confirmation wave from detector $D_{2}$, the transaction forms along the path from $L$ to $D_{1}$ via $A$ and $B$.  The result of the transaction is that a photon is always transferred from the source $L$ to detector $D_{1}$ and that no photons can be transferred to $D_{2}$.  Note that the transaction forms along {\it both} paths from $L$ to $D_{1}$.  This is a transactional account of the operation of the Mach-Zender interferometer.

\begin{figure}
\includegraphics[width=9 cm]{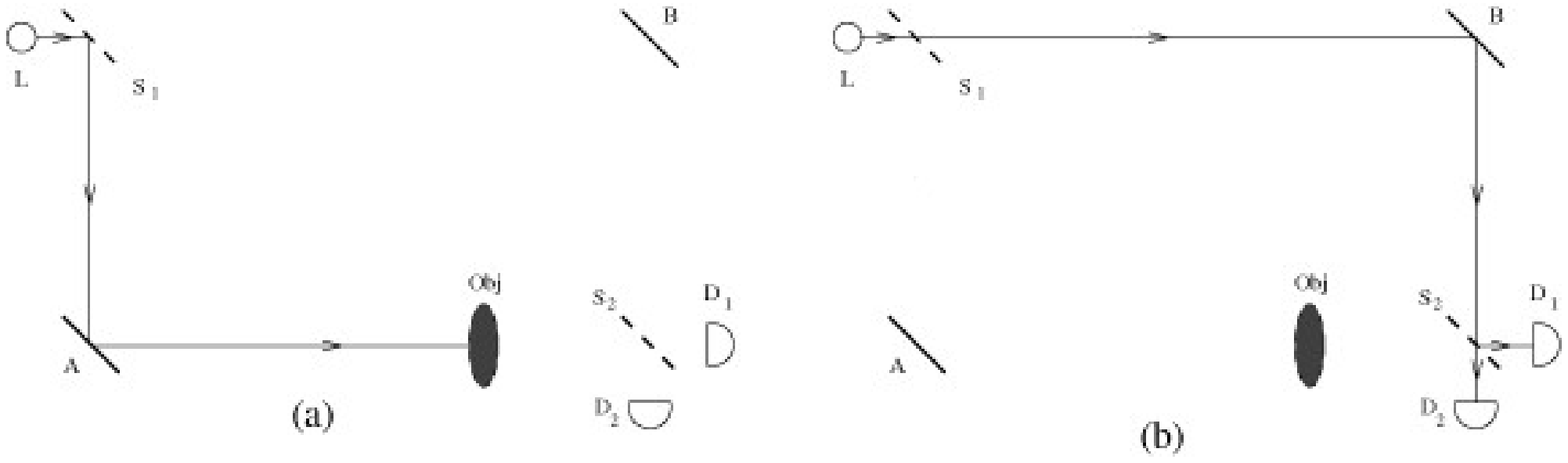}
\caption{\label{four}  (a) Offer waves $\mid L$-$\underline{S_{1}}$-$\underline{A}$-$Obj>$ and (b) $\mid L$-$S_{1}$-$\underline{B}$-$\underline{S_{2}}$-$D_{1}>$ +  $\mid L$-$S_{1}$-$\underline{B}$-$S_{2}$-$D_{2}>$}
\end{figure}

Now let us consider the situation when the object blocks path $A$ as shown in Fig. \ref{four}.  The offer wave on path $A$ is $\mid L$-$\underline{S_{1}}$-$\underline{A}$-$Obj>$.  As before an offer wave on path $B$ is $\mid L$-$S_{1}$-$\underline{B}$-$\underline{S_{2}}$-$D_{1}>$, and it travels from $L$ to detector $D_{1}$. The wave on path $B$ also splits at $S_{2}$ to form offer wave $\mid L$-$S_{1}$-$\underline{B}$-$S_{2}$-$D_{2}>$, which arrives at detector $D_{2}$.  The splits at $S_{1}$ and $S_{2}$ each reduce the wave amplitude by 1/$\sqrt{2}$, so that the offer wave at each detector, having been split twice, has an amplitude of $\frac{1}{2}$.  However, the offer wave $\mid L$-$\underline{S_{1}}$-$\underline{A}$-$Obj>$ to the object in path $A$, having been split only once, is stronger and has amplitude of 1/$\sqrt{2}$.

In this situation, the source $L$ will receive confirmation waves from both detectors and also from the object.  These, respectively, will be confirmation waves $<D_{1}$-$\underline{S_{2}}$-$\underline{B}$-$S_{1}$-$L\mid$, $<D_{2}$-$S_{2}$-$\underline{B}$-$S_{1}$-$L\mid$ and $<Obj$-$\underline{A}$-$\underline{S_{1}}$-$L\mid$.  The first two confirmation waves started from their detectors with amplitudes of $\frac{1}{2}$ (the final amplitude of their respective offer waves) and have subsequently been split twice.  Therefore, they arrive at source $L$ with amplitudes of $\frac{1}{4}$.  On the other hand, the confirmation wave from the object initially has amplitude $1/\sqrt{2}$, and it has been split only once, so it arrives at the source with amplitude $\frac{1}{2}$.

The source $L$ has one photon to emit and three confirmations to choose from, with round-trip amplitudes ($\psi\psi*$) of $\frac{1}{4}$ to $D_{1}$, $\frac{1}{4}$ to $D_{2}$, and $\frac{1}{2}$ to the object.  In keeping with the probability assumption of the transactional interpretation and Born's probability law, it will choose with a probability proportional to these amplitudes.  Therefore, the emitted photon goes to $D_{1}$ 25\% of the time, to $D_{2}$ 25\% of the time, and to the object in path $A$ 50\% of the time.  As we have seen above, the presence of the object in path $A$ modifies the detection probabilities so that detector $D_{2}$ will receive $\frac{1}{4}$ of the emitted photons, rather than none of them, as it would do if the object were absent.

We note that there are also ``aborted" confirmation waves $<D_{1}$-$S_{2}$-$Obj\mid$ and $<D_{2}$-$S_{2}$-$Obj\mid$ from $D_{1}$ and $D_{2}$ along path $A$, which are intercepted by the back side of the interposed object.  They cannot form a transaction because they cannot connect with the source $L$.

How can the transfer of non-classical knowledge be understood in terms of the transactional account of the process?  In the case where there is an object in the $A$ path, it is probed both by the offer wave from $L$ and by the aborted confirmation waves from $D_{1}$ and $D_{2}$.  When we detect a photon at $D_{1}$, (i.e., when a transaction forms between $L$ and $D_{1}$), the object has not interacted with a photon (i.e., a transaction has not formed between $L$ and the object $Obj$).  However, it has been probed by offer and confirmation waves from both sides, which ``feel" its presence and modify the interference balance at the detectors.  Thus, the transactional interpretation gives a simple explanation of the mystery of interaction-free measurements.
	
\section*{Efficient Interaction-Free Measurement with the Quantum Zeno Effect}

\begin{figure}
\includegraphics[width=8 cm]{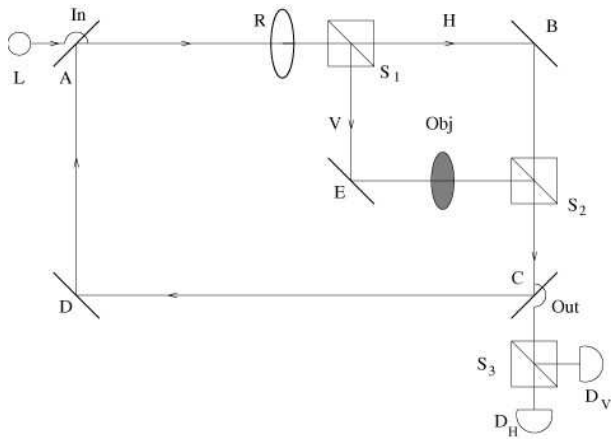}
\caption{\label{five}  Quantum Zeno arrangement for high efficiency interaction-free measurements}
\end{figure}

P. G. Kwiat, et al\cite{kwiat-99}, a collaboration based at LANL and Innsbruck, have demonstrated both theoretically and experimentally that the efficiency of an interaction-free measurement can be increased from 33\% in the EV scheme to a value that is significantly larger.  In fact, the efficiency can be made to approach 100\%, depending on how many times N it is possible to cycle the incident photon through the measurement apparatus.  Their scheme is shown in Fig. \ref{five}.

Here a light source $L$ supplies photons that are horizontally ($H$) polarized.  These are injected ($In$) into an optical ``racetrack" that is capable of cycling a photon around in a closed rectangular loop $N$ times before extracting it ($Out$) to an analyzing system.  After injection, the photon passes through an optical polarization rotator element ($R$) that changes its direction of linear polarization by an angle $\theta = \pi/2N$.  Note that if $N$ is large, this rotation is small.

The photon then travels to a polarization beam splitter ($S_{1}$) that transmit horizontally polarized ($H$) light and reflects vertically polarized ($V$) light.  The object ($Obj$) to be measured may (or may not) be placed in the $V$ beam path.  Downstream of the object position, the $H$ and $V$ photon components enter a second polarization beam splitter ($S_{2}$) that recombines them into a single beam.  The recombined photon then cycles back through the apparatus.  After $N$ cycles, the photon is extracted and sent to a third polarization beam splitter ($S_{3}$) that, depending on its polarization, routes it to a pair of photon detectors ($D_H$ and $D_V$).  This detection is, in effect, a measurement of whether the photon's final polarization is horizontal or vertical.

If no object is in the $V$ path, the polarization split and recombination has no net effect.  The polarization rotator rotates the plane of polarization $N$ times, each time by an angle of $\pi/2N$.  The cumulative rotation is therefore a rotation of $\pi/2$.  Therefore, a photon that was initially polarized horizontally ($H$) will emerge from the apparatus with vertical ($V$) polarization and will be detected by photon detector $D_V$.

On the other hand, if an object is placed in the $V$ path, the $H$ and $V$ beams are not recombined, so the split at the first polarization beam splitter ($S_{1}$) is in effect a polarization measurement.  From Malus' Law, there is a probability $P_{H} = \cos^{2}(\pi/2N)$ that the photon will survive each such horizontal polarization measurement and emerge in a pure state of horizontal ($H$) polarization.   After each cycle in which the photon survives, it is reset to its initial state of horizontal ($H$) polarization, so that when it is extracted after $N$ cycles it and will be detected by photon detector $D_H$.  In each cycle, there is a small probability $(1 - P_{H})$ that the photon will be projected into a state of pure vertical ($V$) polarization, will travel on the $V$ path, will interact with the object, and will be removed from the process.

In summary, if the object is not present, the emerging photon will be detected by the $D_{V}$ detector 100\% of the time.  If the object is present, the emerging photon will be detected by the $D_{H}$ detector with a probability $P_{D} = P_{H}^{N} = \cos^{2N}(\pi/2N)$, and the photon will interact with the object and be removed with a probability $P_{R} = 1 - P_{H}^{N} = 1 - \cos^{2N}(\pi/2N)$.  We note that when $N$ is large, $P_{D}\approx 1 - (\pi/2)^{2}/N$ and $P_{R}\approx (\pi/2)^{2}/N$.  Therefore, the probability of removal decreases as $1/N$ and goes to zero as $N$ goes to infinity. Therefore, the procedure greatly improves the efficiency of interaction-free measurements.  For example, when the number of passes $N$ is equal to 5 the measurement is 60\% efficient.  With $N=10$ it is 78\% efficient, and with $N=20$ it is 88\% efficient.

\section*{A Transactional Interpretation of Quantum Zeno Interaction-Free Measurement}

\begin{figure}
\includegraphics[width=9 cm]{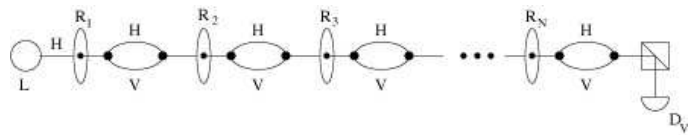}
\caption{\label{six}  Unfolding of the Quantum Zeno measurement (no object) for high efficiency interaction-free measurements}
\end{figure}

Fig. \ref{six} shows an unfolding of the quantum-Zeno interaction-free measurement, for the case where no object is placed in the $V$ beam.

The recycled path is represented as a linear sequence of incremental rotations, beam splittings, and beam recombinations.   It should be clear from the diagram that the successive splittings and recombinations have no net effect.   On the other hand, the $N$ successive rotations have the cumulative effect of a $\pi/2$ rotation that converts the initial horizontally polarized photons into vertically polarized photons by the time they reach the final beam splitter and the detector $D_{V}$.

\begin{figure}
\includegraphics[width=9 cm]{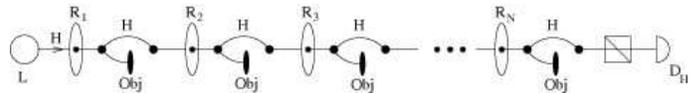}
\caption{\label{seven} Unfolding of the Quantum Zeno measurement with an object present.}
\end{figure}

From the point of view of the transactional interpretation, the initial offer wave leaves the light source $L$ and is then successively rotated, split, and recombined.  These operations do not reduce the amplitude, so the offer wave reaches detector $D_{V}$ at full strength.  The confirmation wave from $D_{V}$ travels back along the same path and arrives back at $L$ at full strength, thereby completing the transaction.
The situation when an absorber is present and there is no interaction is shown in Fig. \ref{seven}.

Now the object $Obj$ blocks the path $V$ of the vertically polarized beam after the splitter, so only the photons on the $H$ path can reach the detector system.  The net effect of this is that after each incremental rotation, the beam is reset to the $H$ state and passes straight through the final splitter without deflection to reach detector $D_{H}$.

From the point of view of the transactional interpretation, the initial offer wave leaves the light source $L$ and at each rotation and splitting the intensity of the offer wave that will reach detector $D_{H}$ is reduced by $\cos(\pi/2N)$, so that the net intensity at the detector is $\cos^{N}(\pi/2N)$.  At the m$^{th}$ split (for m=1 to N), an offer wave of intensity $\cos^{m-1}(\pi/2N) \sin(\pi/2N)$ travels to the object $Obj$ and may interact with it.  The confirmation wave from each of these potential interactions will travel back to the light source $L$ with the same reduction factor, so that the net probability of an interaction following the $m$th split is $\cos^{2(m-1)}(\pi/2N) \sin^{2}(\pi/2N)$.

\begin{figure}
\includegraphics[width=9 cm]{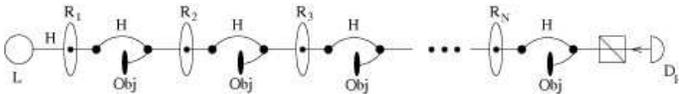}
\caption{\label{eight} Paths of detector confirmation waves in the unfolded Quantum Zeno measurement with an object present.}
\end{figure}

The path of offer waves from the detector $D_{H}$ is shown in Fig. \ref{eight}.

The confirmation wave leaves detector $D_{H}$ with an amplitude of $\cos^{N}(\pi/2N)$, the final amplitude of the offer wave.  As the confirmation wave travels back to the light source $L$, at each of the $N$ splits it is reduced in intensity by a factor of $\cos(\pi/2N)$.  Thus its net intensity at $L$ will be $\cos^{2N}(\pi/2N)$, which is just the probability that the detection event will occur.  At each split, there is a component of the confirmation wave that takes the lower path in Fig. \ref{eight} and ends at object $Obj$.  However, these components cannot form a transaction, since they cannot connect back to the light source $L$.

As before, the object $Obj$ in the $V$ path is probed both by the offer wave from $L$ and by the aborted confirmation wave from $D_{H}$.  When we detect a photon at $D_{H}$, (i.e., when a transaction forms between $L$ and $D_{H}$), the object $Obj$ has not interacted with a photon (i.e., a transaction has not formed between $L$ and $Obj$), but the object has been probed repeatedly by weak offer and confirmation waves from both sides.  As the number of passes $N$ is increased and the efficiency of the measurement approaches 100\%, the amplitudes of these probe waves grows weaker as their number increases.
It also becomes clear why, even when the object does not interact with a photon, the {\it possibility} of interactions is required.  If the interaction probability were zero, the offer and confirmation waves would not be blocked by the interposed object and the measurement would not have been possible.

\section*{Conclusion}

We have found that the transactional interpretation of quantum mechanics has provided a straightforward and economical account of interaction free measurements of the Elitzur-Vaidmann type.  It has explained the source of the non-classical information that is obtained from such interaction-free measurements.  It also provides some support for the transactional interpretation, since offer and confirmation waves that are not a part of a transaction play key roles in the measurement process.

We have also shown how, in two rather complicated experiments in quantum optics the transactional interpretation can be used to visualize the processes involved, even when the outcome leads to counter intuitive and non-classical results.  This demonstrates the power of the transactional interpretation as a tool for experimental design and analysis.


\begin{thebibliography}{99}

\bibitem{cramer-86} J. G. Cramer, Reviews of Modern Physics 58, 647-687 (1986).
\bibitem{elitzur-93} A. C. Elitzur and L. Vaidman, Foundations of Physics 23, 987-997 (1993).
\bibitem{mandel-95} L. Mandel and E. Wolf, Optical Coherence and Quantum Optics, Cambridge University Press (1995).
\bibitem{kwiat-99} P. G. Kwiat, et al., Phys. Rev. Letters 83 4725-4728 (1999).

\end{thebibliography}
\end{document}